\documentclass[12pt,preprint]{emulateapj}  

\shorttitle{{\small CGCG 480-022}: A DISTANT LONESOME MERGER?}
\shortauthors{CARRETERO ET AL.}

\begin{document}

\title{{\small CGCG 480-022}: A distant lonesome merger?}

\author{C. Carretero\altaffilmark{1}, A. Vazdekis\altaffilmark{1}, A. C. Gonz\'alez-Garc\'{i}a\altaffilmark{1}, J. E. Beckman\altaffilmark{1,}\altaffilmark{2} and V. Quilis\altaffilmark{3}} 
\altaffiltext{1}{Instituto de Astrof\'{\i}sica de Canarias, V\'{\i}a L\'actea s/n, 38200 La Laguna, Tenerife, Spain; cch@iac.es, vazdekis@iac.es, cglez@iac.es} 
\altaffiltext{2}{Consejo Superior de Investigaciones Cient\'{\i}ficas, Spain; jeb@iac.es}
\altaffiltext{3}{Departamento de Astronom\'{i}a y Astrof\'{i}sica, Universidad de Valencia, E-46100 Burjassot, Valencia, Spain; vicent.quilis@uv.es}

\begin{abstract}
We present a complete analysis, which includes morphology, kinematics, stellar populations and N-body simulations of {\small CGCG 480-022}, the most distant (c$z=14317~{\rm km~s}^{-1}$) isolated galaxy studied so far in such detail. The results all support the hypothesis that this galaxy has suffered a major merger event with a companion of $\sim0.1$ times its mass. Morphology reveals the presence of a circumnuclear ring and possibly further ring debris. The radial velocity curve looks symmetrical, whilst the velocity dispersion increases with radius reaching values that do not correspond to a virialized system. Moreover, this galaxy deviates significantly  from the Fundamental Plane and the Faber-Jackson relation. The stellar population analysis show that the ring is younger and more metal-rich, which suggest that it has undergone a fairly recent burst of star formation. Both morphological and dynamical results are in broad agreeement with our N-body simulations.
\end{abstract}


\section{Introduction}

Elliptical galaxies located in very low density environments are highly uncommon but hold important clues about galaxy formation and evolution. For example, current hierarchical models for galaxy formation predict that ellipticals in low density environments have stellar populations which are younger by 1-2~Gyr than their cluster counterparts \citep[e.g.][]{baugh98,clemensetal06,delucia06}. In this framework, isolated ellipticals are thought to be formed in mergers of galaxy pairs or small groups of galaxies \citep{jones00,donghia05}. In the first case, the mergers will give rise to kinematical misalignments whilst, in the second, they will produce an increasing amplitude of rotation with galactocentric radius, and also rotation around the minor axis \citep{weilhernquist96}.

Detailed studies involving morphology, kinematics and/or stellar populations of isolated ellipticals have been restricted to nearby galaxies \citep[c$z$$<$10000 km s$^{-1}$; e.g.][]{collobertetal06}. In this paper, we perform an exhaustive analysis of {\small CGCG 480-022} (c$z$=14317 km s$^{-1}$), which includes morphological, kinematical, stellar population and N-body simulation properties. We present evidence supporting the hypothesis that this galaxy suffered a major merger event.

We adopt a cosmological model with H$_0$=71 km s$^{-1}$ Mpc$^{-1}$, $\Omega_m$=0.3 and $\Omega_\Lambda$=0.7.

\section{Observations and Data Reduction}

Optical and near-IR images were extracted from the DSS and 2MASS catalogues, respectively.

Spectroscopy was obtained with the DOLORES spectrograph on the 3.5m TNG telescope at La Palma (Spain) on October 3, 2005. We used the medium-resolution MR-B\#2 grism, in the wavelength range $\lambda\lambda$3500-7000~\AA, and a 1.1$''$$\times$50$''$ slit. The dispersion was 1.7~\AA/pix with an instrumental resolution of 8 \AA\ (FWHM). We obtained 10$\times$30 minute exposures, interspersed with He+Ar arc-lamp spectra, flux and radial velocity standards in the same instrumental setup. The seeing  was 1.1$''$.

Data reduction was performed using standard IRAF packages. Once wavelength and flux calibrated, the spectra were extracted in the spatial direction coadding 4 pixels to reach the seeing size, or more if necessary to reach S/N$\ge$30 \citep[see][]{cardiel98}. Following this procedure, we obtained 13 bins out to $r$=$\pm$0.8~R$_{\rm eff}$.

\section{Analysis and Results}

\subsection{Morphology}

\begin{figure}
\epsscale{0.6}
\plotone{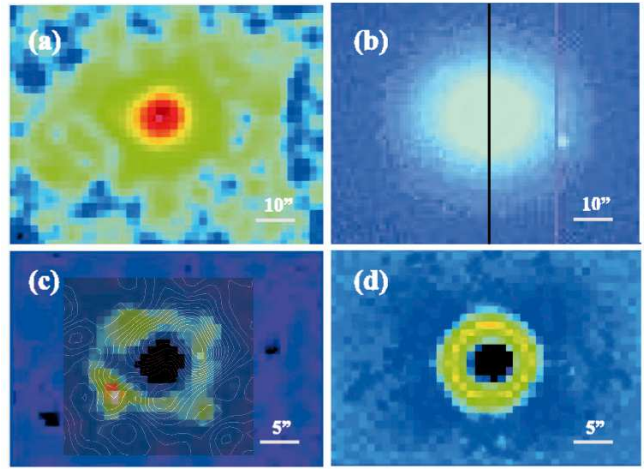}
\caption{Morphology analysis. Panels (a) and (b) show the actual images of {\small CGCG 480-022} in the K-band and optical, respectively. Panels (c) and (d) present the residuals when subtracting a 2D model of the galaxy from the images using GALFIT. In both cases, a ring of $r$$\sim$5~$arcsec$ is clearly visible. It is noteworthy that some tail-like structures appear in the outer regions of panel (a). Contours in panel (c) correspond to b-r color values. The spectrograph slit is overplotted on panel (b).}
\label{morfologia}
\end{figure}

To determine morphological and photometric parameters we used the code described in \citet{trujillo01} on K-band and optical images. Table \ref{table1} presents, for each morphological parameter, the average value obtained from both bands. Figure \ref{morfologia}a shows the K-band image while Figure \ref{morfologia}b shows the optical image with the spectrograph slit overplotted. According to our morphological parameters, and its optical appearance, {\small CGCG 480-022} seems to be a typical elliptical galaxy. However, when inspecting the K-band image, we detect possible tidal tails or ring debris. In order to search for the existence of more subestructures, we performed a 2D fit of the galaxy using {\small GALFIT} \citep{peng02} with those parameters listed in Table \ref{table1}. Figures \ref{morfologia}c and \ref{morfologia}d show the residuals obtained when subtracting 2D models from each image in K-band and optical, respectively. In both cases we clearly observe a well-defined ring at $r$$\sim$5~$arcsec$, yielding to a rippled structure similar to the shells found in other galaxies \citep[e.g.][]{quinn84}. Residuals in K-band are compatible with b-r color contours overplotted on Figure \ref{morfologia}c.

\subsection{Kinematics}

\begin{figure}
\epsscale{0.4} 
\plotone{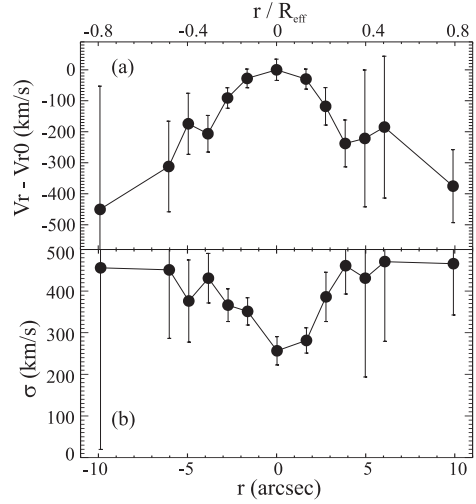}
\caption{Kinematical results. {\em Top}: Radial velocity relative to the central value versus radius. Note that the symmetric shape does not correspond to a typical rotation curve. {\em Middle}: Velocity dispersion versus radius.}
\label{dinamica}
\end{figure}

The radial velocity for each bin was determined by cross-correlating the observed spectrum with a set of representative synthetic single stellar population spectra, from the model of \citet{vaz99}. This method is described in \citet{bottema88}, following the paper of \citet{tonrydavis79}. The measured $V_{\rm r}$ values are represented versus radius in Figure \ref{dinamica}a. Surprisingly, $V_{\rm r}$ decreases with galactocentric distance at both sides of the centre, out of $250~{\rm km~s}^{-1}$ at $r$$\sim$3~$arcsec$. We also see a relative increase of $100~{\rm km~s}^{-1}$ where the ring is located. Note that the plotted region represents 0.8$R_{\rm eff}$.

To measure the velocity dispersion, $\sigma$, we followed the method described in \citet{davies93}. Figure \ref{dinamica}b shows $\sigma$ vs. radius. We observe that, within the errors, regions outside the ring present an almost constant $\sigma$$\sim$400~km~s$^{-1}$ while the inner parts look more ordered, i.e. $\sigma$ decreases, reaching $250~{\rm km~s}^{-1}$ in the centre. A similar profile has been found for the isolated massive galaxy {\small NGC 1700} \citep{bergmann02}.

\subsection{N-body Simulations}

To aid our understanding of these results we have performed N-body simulations. Our numerical models simulated a merger between two elliptical galaxies with a mass ratio of 10:1 and 5:1, and the fly-by of a dwarf galaxy about a massive elliptical. As initial conditions for our experiments we used: {\em i)} isotropic spherical \citet{jaffe83} models for the elliptical galaxies, and {\em ii)} a \citet{king66} model for the dwarf galaxy, with a halo modeled with an Evans model. See \citet{gonzalezgarcia05} for further details of this model and initial conditions for the elliptical galaxies, and \citet{kuijken95} for the dwarf galaxy model.

We used the tree-code {\small GADGET1.1} \citep{springel01} employing 10$^5$ particles for each model. Low mass Jaffe models are scaled down homologous systems as the model with mass 1 and with 10 or 5 times fewer particles. Parabolic grazing orbits are employed for the merger simulations while a hyperbolic encounter is used for the fly-by experiment. Energy was well conserved, with errors well below 0.1\%.

The material originally belonging to the small galaxy in the 5:1 merger finally lies on a fat disk-like distribution, while that of the 10:1 merger produces a torus-like body. The fly-by encounter produces shells and ripples in the inner parts of the giant galaxy from particles belonging to the satellite; however no ring or disk is produced in this encounter.

\begin{figure}
\epsscale{0.7} 
\plotone{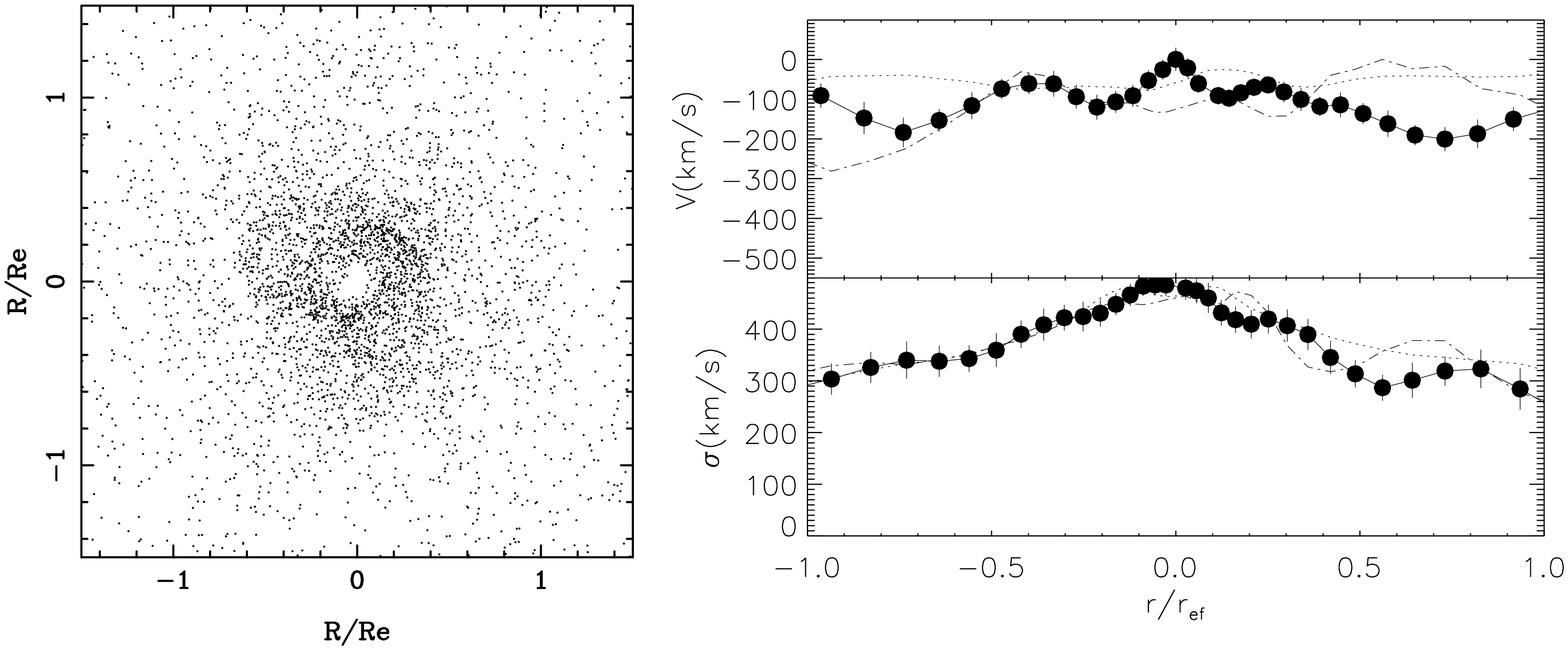}
\caption{N-body results. {\em Left}: Material from the smaller system in the 10:1 merger remnant (for clarity, the large galaxy component is not shown). Note the presence of a well-defined ring at a galactocentric radius similar to that of the observed ring. {\em Right}: Kinematics for all the visible material of the merged system. Solid, dashed-dotted and dotted lines correspond to the system at 2, 2.1 and 5~Gyr after the merger, respectively. {\em Right-top}: Radial velocity relative to the central value versus radius. {\em Right-bottom}: Velocity dispersion versus radius.}
\label{nbody}
\end{figure}

We have obtained a broad estimate of the mass ratio between our target galaxy and a possible companion. If we consider the merger as an inelastic collision, then the whole kinetic energy of the satellite will be transformed into `thermal' energy of the stars of the main galaxy. This process will create a velocity dispersion, $\sigma$, given by $\frac{1}{2} m v_{sat}^2$=$\frac{1}{2} M \sigma^2$$\Rightarrow$$\frac{M}{m}$=$\left(\frac{v_{sat}}{\sigma}\right)^2$, where v$_{sat}$ is the relative velocity of the satellite with respect to the main galaxy, and $M$ ($m$) the mass of the central (satellite) galaxy. Considering that they are field galaxies, the proper motion should be of the order of $v_{sat}$$\sim$10$^3$~km~s$^{-1}$, while $\sigma$$\sim$300~km~s$^{-1}$. Then, we find $\frac{M}{m}$$\sim$10. Moreover, of the models explored, the E+E 10:1 merger is the one with properties closer to those reported for {\small CGCG 480-022}. Figure \ref{nbody} presents the particle distribution of the 10:1 system. A ring is clearly observable at a galactocentric distance comparable to that detected in {\small CGCG 480-022}. The global kinematics as seen from the same point of view are given in Figure \ref{nbody}. The velocity curve presents a drop as we move to larger radii, in agreement with the observations, but the velocity dispersion presents a general declining profile, contrary to what is observed.

\subsection{Stellar Populations}

The stellar population analysis was performed following the prescriptions described in \citet{carretero04}. To derive mean luminosity-weighted ages and metallicities, we compared selected absorption line strengths with those predicted by the model of \citet{vaz99}. This model provides flux-calibrated spectra in the optical range at a resolution of 1.8~\AA\ (FWHM) for single-burst stellar populations. This way, we can transform synthetic spectra to the resolution and dispersion of the galaxy spectrum. Plots of the strengths of selected indices \citep[Mg$_2$ and Fe4383;][]{wor94a} versus H$\beta$ provide close to orthogonal model grids, allowing an accurate estimation of galaxy mean age and abundances of the different elements. We have determined mean ages and relative abundances [Mg/H] and [Fe/H] at each galactocentric distance, computing their radial gradients.

Figure \ref{poblaciones} presents the ages and abundances for each element with respect to the central values, versus the radial distance. If we exclude those regions within 3$<$$|r|$$<$6~$arcsec$, we observe clear gradients for both ages and abundances, with the age increasing towards outer regions and abundances decreasing throughout the galactic radius. This result is in agreement with several studies involving gradients in elliptical galaxies \citep[e.g.][]{davies93,fisher95,kobayashiarimoto99}.

When considering the region 3$<$$|r|$$<$6~$arcsec$, where the ring is located, we observe a decrease in age with respect to the surrounding regions, associated with a bump in metallicity. This suggests the occurrence of a star formation burst coupled with the presence of the ring.

\section{Discussion}

The morphological, kinematical, N-body simulations and stellar population analysis all suggest that {\small CGCG 480-022} could have experienced a major merger event with an alleged companion. In this section, we will focus on the evidences provided by each approach.

{\em Morphology}. The residuals obtained when substracting a 2D model from the images show, in all bands, the presence of a ring of $r$$\sim$5~$arcsec$. This ring, together with outer streams of material appearing in K-band images, could trace the propagating sound-waves produced by a merger \citep[e.g.][]{hernquist87} more likely than by the action of tidal interactions or asymmetric star formation\footnote{Even though we found a star formation burst in the ring, which would be induced by the passing wave; i.e. the varying SF history should be the effect of the ring structure, not the cause.} \citep{colbert01}. In fact, \citet{malincarter83} found that morphological peculiarities, such as shells or ripples, occur roughly 5 times more frequently in environments outside of rich clusters. This result was confirmed by \citet{colbert01} who concluded that shells are much more prevalent in isolated galaxies than in group galaxies. Note that the ROSAT X-ray image of the galaxy \citep{zimmermann01} shows two symmetric lobes in the N-S direction.

{\em Kinematics}. The most intriguing of our results concerns the kinematics of the galaxy. First, within the errors, we find a constant $\sigma$$\sim$400~km~s$^{-1}$ which decreases in the interior of the ring. A similar pattern has been found in other isolated massive galaxy \citep[{\small NGC1700};][]{bergmann02}. However, such high $\sigma$ does not correspond to a virialized system but, in a merger scenario, the intrinsic velocity dispersion should be affected by the collision, yielding higher values of $\sigma$. Second, the radial velocity curve looks symmetrical. Furthermore, both $\sigma$ and $V_{\rm r}$ values bump at the position where the ring is located, i.e. 3$<$$|r|$$<$6~$arcsec$. As an additional test, we checked whether {\small CGCG 480-022} follows or not the Fundamental Plane \citep[FP;][]{djorgovskidavis87} and the Faber-Jackson \citep{faberjackson76}, the Kormendy \citep{kormendy77} and the Mg$_2$-$\sigma$ \citep{terlevich81} relations. We considered these relations in the near-IR, with data taken from \citet{pahre98}. {\small CGCG 480-022} deviates $3\sigma$ from the FP, $6\sigma$ from the Faber-Jackson relation and $4\sigma$ from the Mg$_2$-$\sigma$ relation; but just $0.2\sigma$ from the Kormendy relation. Our galaxy is an outlier in those relations involving the velocity dispersion, in the sense that it {\em should} be lower {\em if} it corresponds to the intrinsic velocity dispersion of a virialized system. Thus, a major event affecting the dynamics of this galaxy must have occurred. A similar explanation could be proposed for the $V_{\rm r}$ curve, which does {\em not} correspond to a rotating system. In order to disentangle the kinematical peculiarities found in this galaxy, future 2D-spectroscopy  will be extremely helpful, specially to check if our derived kinematics depend on the slit direction.

{\em N-body simulations}. The N-body experiments performed suggest that a possible formation mechanism is a merger between two elliptical galaxies with a mass ratio near 10:1, and meeting each other on a grazing orbit. In such an encounter the smaller system spirals inwards while the merger is proceeding finally into a torus-like structure. Such torus as seen face-on resembles a ring and the kinematics imposed to the overall field would present a discontinuity. A significant success for this model is that it reproduces without fine tuning the velocity profile observed  in {\small CGCG 480-022}. The central zones moves with a larger velocity than the zones at increasing galactocentric distances, with respect to the systemic velocity of  the galaxy, and there are secondary maxima at the radial location of the ring in both directions from the center. The range of values of the velocity dispersion predicted by the model is commensurate with that observed, but the decline in value as the center is approached is not reproduced. A reasonable explanation for this is that more ordered motion near the galaxy center could result from an inner disk within the ring radius. The present merger simulation has not, for simplicity, included the presence of  gas, but merger models which do include gas \citep[e.g.][]{barnes02} show a tendency for the gas to fall towards the center of the galaxy and form a disk. Also, more detailed N-body+SPH simulations including star formation and feedback processes will help to better understand the central kinematics of the system. 

\begin{figure}
\epsscale{0.4} 
\plotone{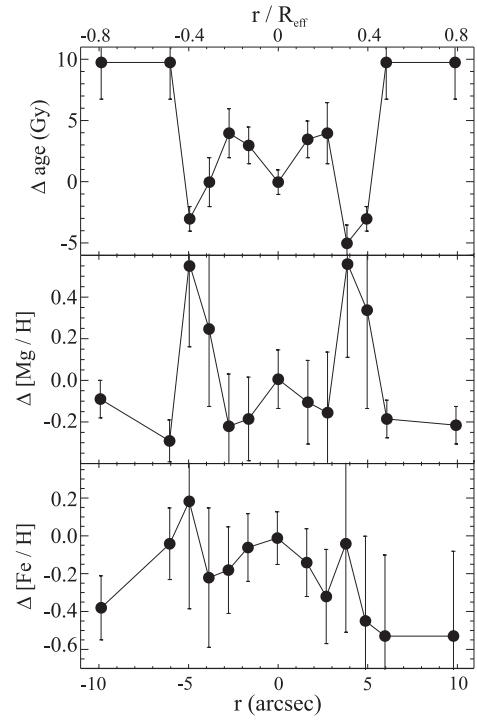} 
\caption{Results of the stellar population analysis. The plots show the values of the estimated ages and abundances of individual elements, relative to the centre. Expected radial gradients for both ages and metallicities are obtained if we exclude the region 3$<$$|r|$$<$6~$arcsec$. At radial distances corresponding to the position of the ring, stellar population analysis indicates the presence of a recent star formation burst, with younger and more-metal rich stars.} 
\label{poblaciones}
\end{figure}

{\em Stellar populations}. Following our merger hypothesis, there should be a star formation burst in the ring due to the passing sound-wave. Therefore, the stars in the ring should be younger and more metal-rich than those within inner regions, as we found. Given that we know the geometry of the ring, we can broadly estimate the crossing time of the sound-wave through the position of the ring, under the following assumptions: {\em i)} the sound speed is constant throughout the whole system, and {\em ii)} the system behaves as an ideal gas. Then, the mean pressure of a system of particles with a given velocity dispersion, $\sigma$, is $p$=$\frac{1}{3}\rho\sigma^2$. The propagation velocity of the perturbation corresponds to the sound speed, $c_s$, as $c_s^2$=$\frac{{\rm d}p}{{\rm d}\rho}$=$\frac{1}{3}\sigma^2$$\Rightarrow$$c_s$$\sim$180~km~s$^{-1}$. Considering that the ring radius is $\sim$5~$arcsec$, then the ellapsed time since the formation of the perturbation is $t(s)$=4.8$\times$10$^{-6}D/c_s$, where $D$ is the distance to the galaxy in km. Therefore, $t$ should be the stellar age difference between the ring and the centre. Given that $D$=6.2$\times$10$^{23}$~km, then $t$=5.3 Gyr, which is in good agreement with our stellar population results (see Figure \ref{poblaciones}). Even though approximative, these numbers do support the merger hypothesis.\\

A question we would like to pose is why the most distant isolated elliptical galaxy so far studied in such detail looks so peculiar. To confirm the results presented here and probe the origin of this system, deep 2D spectroscopy is strongly needed. We also note that {\small CGCG 480-022} is not a completely isolated galaxy but has one confirmed companion at a distance of 0.25~Mpc: {\small PGC 003298} \citep{beers83}. The DSS image of this galaxy shows an elongated shape in the direction of {\small CGCG 480-022}, whilst J- and K-band 2MASS images present some tail-like structures. Will {\small PGC 003298} be the next to merge?

\acknowledgements
The authors thank S. Patiri, I. Trujillo, D. Bettoni and the referee for useful comments. AV and VQ are Ram\'on y Cajal Fellows of the Spanish Ministry of Education and Science. This work has been supported by the Spanish Ministry of Education and Science (grants AYA2004-03059 and AYA2004-08251-C02-01) and the Generalitat Valenciana (grant GV2005/244).


\begin{deluxetable}{ccccccc}
\tablecolumns{7} 
\tablecaption{Determined properties of {\small CGCG 480-022}} 
\tablehead{ c$z$ & m$_{\rm k}$ & $\langle\mu_{\rm k}\rangle$ & $R_{\rm eff}$ & $n$ & $e$ & P.A.\\
  km/s & mag & mag/$''^2$ & $arcsec$ & & & deg} 
\startdata  14317 & 11.01 & 19.1 & 13.0 & 3.87 & 0.9 & 89
\enddata
\label{table1}
\end{deluxetable}


\end{document}